\documentclass[aps,prd,onecolumn,amssymb]{revtex4}
\usepackage{graphicx,bm,color}
\usepackage{amsmath}
\usepackage{amssymb}
\usepackage{amsfonts}
\usepackage{epsfig}
\usepackage{float}
\usepackage{graphicx}

\newcommand{\be}{\begin{equation}}
\newcommand{\ee}{\end{equation}}
\newcommand{\bea}{\begin{eqnarray}}
\newcommand{\eea}{\end{eqnarray}}
\newcommand{\beaa}{\begin{eqnarray*}}
\newcommand{\eeaa}{\end{eqnarray*}}

\allowdisplaybreaks[4]
\begin{document}

\title{Gravitational memory effects of black bounces and a traversable wormhole}
\author{Hamed Hadi$^{1,2}$}\email{hamedhadi1388@gmail.com}
\author{Reza Naderi$^2$}\email{r_naderi8@yahoo.com}
\affiliation{$^1$Faculty of Physics, University of Tabriz, Tabriz 51666-16471, Iran\\$^2$Department of Physics, Azarbaijan Shahid Madani University, Tabriz, 53714-161 Iran
}

\begin{abstract}
	Black bounces are spacetimes that can be interpreted as either black holes or wormholes depending on specific parameters. In this study, we examine the Simpson-Visser and Bardeen-type solutions as black bounces and investigate the gravitational wave in the background of these solutions. We then explore the displacement and velocity memory effects by analyzing the deviation of two neighboring geodesics and their derivatives influenced by the magnetic charge parameter $a$. This investigation aims to trace the magnetic charge in the gravitational memory effect. Additionally, we consider another family of traversable wormhole solutions obtained from non-exotic matter sources to trace the electric charge $Q_{e}$ in the gravitational memory effect, which can be determined from the far field asymptotic. Furthermore, this research aims to explore the gravitational memory effect related to the variation in Bondi mass for the Simpson-Visser and Bardeen-Type black bounces. The investigation will also be conducted on a traversable wormhole solution that does not require any exotic field. This study holds importance in not only identifying compact objects such as wormholes through gravitational memory effects but also in observing the charge $Q_{e}$, which offers a tangible manifestation of Wheeler's idea of "electric charge without charge."
\end{abstract}
\maketitle

\section{Introduction}
General relativity (GR) is widely acknowledged as the most suitable and comprehensive theory for elucidating the gravitational interaction \cite{a1,a2}. This theory has exhibited remarkable success in addressing numerous quandaries and possesses the capacity to predict novel phenomena \cite{a3,a4,a5,a6,a7,a8,a9,10}. Notably, the equations formulated by Einstein yield intriguing solutions, such as the Schwarzschild black hole. This particular black hole is distinguished by its static and symmetrical nature, devoid of both spin and charge \cite{11}. Furthermore, GR presents solutions known as wormholes, which establish connections between disparate points in space-time within the same universe or even across different universes via a tunnel \cite{13}. The characteristics of these wormhole solutions, including their traversability, resemblance to black holes, and stability, have been extensively investigated in various scholarly sources \cite{14,15,16,17,18,19,20,21,22,23,24,25,26,27,28,29,30,31,32,33,34,35,36,37}.

Furthermore, Einstein equations have been shown to have solutions for black holes and wormholes through the use of a nonlinear electromagnetic source. Additionally, a distinct type of solution known as regular black holes has been discovered, which possess an event horizon but lack a singularity. The original proposal for this solution was made by Bardeen \cite{Bardeen}. One notable feature of regular black holes is the deviation of photons from geodesics, as well as changes in the thermodynamics of these solutions \cite{40,41,42}. For more information on this solution, please refer to \cite{43,44,45,46,47,48,49,50,51,52,53,54,55}.

The black bounce, a novel variant of a regular solution, has the unique ability to transform into either a black hole or a wormhole, depending on the specific parameters chosen. One of the most notable examples of this intriguing solution is the Simpson-Visser black bounce \cite{SV}, which features a throat located at $r=0$ and an event horizon area that remains unaffected by the solution's parameters. For alternative models of black holes, references \cite{57,58,59,60} may be consulted. Extensive investigations have been conducted on various properties of these models, without explicitly attributing them to the source of matter \cite{61,62,63,64,65,66,67,68,69,70,71,72,73,74,75,76,77,78,79}. However, it is important to note that nonlinear electromagnetic fields alone are insufficient in describing the matter source responsible for the existence of the black bounce in the context of GR. To identify a suitable source for this solution, the inclusion of scalar and phantom fields becomes necessary, as discussed in \cite{main}.

One of the challenges in constructing traversable wormholes is the requirement of exotic matter. To circumvent this problem, some alternative approaches based on generalized theories of gravity have been investigated. Among these, higher-curvature theories of gravity provide a promising framework for the existence of stable wormholes. A notable example is the low-energy heterotic string effective theory \cite{b7b7,b8b8}, which generalizes the four-dimensional gravitational theory by adding extra fields and higher-curvature terms to the standard Einstein-Hilbert action. Within this scenario, some studies have explored the possibility of traversable wormholes without the need for any exotic matter, using the dilatonic Einstein-Gauss-Bonnet theory in four-dimensional spacetime \cite{a7a7}. Moreover, recent research has attempted to construct a traversable wormhole solution by employing fermions and linear electromagnetic fields as the source terms \cite{prl}.

This paper investigates the black bounce solutions and traversable wormholes in the context of gravitational memory effects, using fermions and linear electromagnetic fields as the source terms. The gravitational memory effect refers to the imprint of gravitational waves (GW) on the background metric that satisfies Einstein's equations. This phenomenon, which was discovered long ago \cite{m1,m2,m3,m4}, originates from the propagation of energy flux to the future null infinity, where the spacetime becomes asymptotically flat \cite{x1}. The region where this effect takes place is governed by an asymptotic symmetry group called the Bondi-Metzner-Sachs (BMS) group \cite{bondii,bondi1,bondi2}. The study of the infrared structure of gravity is of great importance for physicists, and the memory effect is one of the key features of interest \cite{infrared1,infrared2,infrared3}.

The main focus of this paper is to examine the Bondi-Sachs formalism \cite{bondii,33bb}, which presents a new approach to the study of GWs within the framework of GR. This formalism, proposed in \cite{41bb}, utilizes outgoing null rays that trace the path of the waves in spacetimes with axisymmetry. The primary goal of this research is to detect gravitational memory effects associated with black bounce solutions and a traversable wormhole solution. Bondi-Sachs \cite{41bb,42bb}, has successfully extended this formalism to non-axisymmetric spacetimes and has determined the asymptotic symmetries as one approaches infinity along the outgoing null hypersurfaces.
The study carried out in \cite{34bb}, concentrated on the constraints associated with the three possible origins of gravitational memory effect within the linearized Bondi-Sachs framework. The results indicated that the only recognized source of B-mode gravitational memory is of primordial origin. In the linearized theory, this corresponds to a consistent wave entering from past null infinity.
The correlation between gravitational memory effects and the BMS symmetries of asymptotically flat spacetimes within the scalar-tensor theory was thoroughly examined in a detailed investigation referenced in \cite{35bb}. The study successfully derived solutions to the equations of motion near the future null infinity by employing the generalized Bondi-Sachs coordinates and ensuring compliance with a specific determinant condition. In \cite{36bb}, the study explored the expansion of Bondi-Sachs form to alternative modified gravity theories such as Brans-Dicke theory. The investigation focused on the relationship between memory effects, symmetries, and conserved quantities within Brans-Dicke theory. Additionally, the field equations were computed using Bondi coordinates, and a specific set of boundary conditions were established to represent asymptotically flat solutions within this framework. Consequently, the analysis of the asymptotic symmetry group for these spacetimes revealed its alignment with the BMS group in GR.

In reference \cite{37bb}, the authors revisit the metric derivation within Newman-Unti and Bondi gauges, with a specific focus on the complete quadrupole-quadrupole interactions. They then proceed to rederive the displacement memory effect and present formulas for all relevant Bondi aspects and dressed Bondi aspects, which are essential for analyzing both the primary and secondary memory effects. Furthermore, they successfully calculate the Newman-Penrose charges, the BMS charges, and the celestial charges of second and third order. These charges are established based on the known second order and innovative third order dressed Bondi aspects, taking into account interactions that involve mass monopole-quadrupole and quadrupole-quadrupole. The study of the GW memory is being carried out in the context of a specific class of braneworld wormholes that do not require exotic matter fields for their traversability, as detailed in \cite{38bb}. The memory effect, which indicates the influence of extra dimensions and the wormhole characteristics of spacetime geometry, is elaborated upon in \cite{38bb}. Additionally, it has been proposed in \cite{39bb,40bb} that the charges linked to the internal Lorentz symmetries of GR, in combination with the incorporation of higher derivative boundary terms in the action, can capture detectable GW phenomena.

The detection of GWs, as reported in \cite{GW1,GW2,GW3,GW4,GW5,GW6,GW7,GW8}, reveals another intriguing aspect of the gravitational memory effect. This is another topic that we will concentrate on in this paper. We will investigate the gravitational memory effects by measuring the deviation of two nearby geodesics in this solution due to a GW pulse. We will examine the Simpson-Visser and Bardeen-Type black bounce solutions and a traversable wormhole solution in the presence of fermions and linear electromagnetic fields as the source terms under the influence of a GW pulse, in order to find the signatures of these compact objects in the deviation of two nearby geodesics as memory effects. The parameters of these solutions determine their impact on the deviation of two nearby geodesics.

In this manuscript, we explore the variation in Bondi mass at future null infinity of a system with a geometric background of black bounce or a traversable wormhole in the presence of a GW superimposed on these backgrounds. The alteration in the Bondi mass is a consequence of the GW within the background. The geodesic memory effect for the Simpson-Visser solution and the Bardeen-Type black bounce solution will be examined in sections II and III, respectively. Section IV will delve into the features of the traversable wormhole solution with fermions and linear electromagnetic fields as the source terms, along with their implications for the memory effects. Section V  will be dedicated to investigating the gravitational memory effect concerning the change in Bondi mass for the Simpson-Visser and Bardeen-Type black bounces. This analysis will be reiterated for a traversable wormhole solution that does not involve any exotic field or matter in section VI. Ultimately, our conclusions will be presented in section VII.

\section{Geodesic memory effect for Simpson-Visser solution as a black bounce}
Within this section, we explore the geodesic memory effect by examining the deviation of geodesics between neighboring points. This deviation arises due to the propagation of a GW, with the separation of geodesics measuring the extent of the displacement memory effect. Furthermore, we also investigate the velocity memory effect that occurs following the passage of a GW pulse. Hence, in this particular model, we have a black bounce solution serving as the background, with the GW pulse being introduced to this backdrop. 

Recent studies \cite{1,2,3,4,5} have examined these phenomena by analyzing the trajectory of geodesics in precise plane GW spacetimes. To achieve this, a Gaussian pulse is selected to represent the polarization in the metric of the plane GW spacetime. Subsequently, the geodesic equations are solved using numerical methods, allowing for the calculation of the displacement and velocity memory effects, which correspond to the changes in separation and velocity, respectively. Additionally, alternative theories of gravity have been explored within this framework \cite{6,7}.

In this study, we investigate the development of geodesics within the black bounce background while considering the influence of a GW pulse. We demonstrate the relationship between the displacement and memory effects and their dependence on this particular background. The line element that describes a black bounce can be expressed as the general solution.
\begin{equation}\label{bouncemetric}
	ds^{2}=-f(r)dt^{2}+f(r)^{-1}dr^{2}+\sigma(r)^{2}(d\theta^{2}+\sin^{2}\theta d\phi^{2}).
\end{equation}
If we constrain the solution to Simpson-Visser black bounce then we have
\begin{equation}
	f(r)=1-\frac{2m}{\sqrt{r^{2}+a^2}}, ~~~and~~~ \sigma(r)=\sqrt{r^{2}+a^2}.
\end{equation}
The magnetic charge, denoted by the parameter $a$, assumes a significant role in averting the emergence of a singularity at $r=0$. This fact can be substantiated through the computation of the \textit{Kretschmann scalar}, a scalar invariant that quantifies the curvature of spacetime. The expression for the \textit{Kretschmann scalar} corresponding to this particular solution is provided as follows.
\begin{equation}
K=R_{\mu\nu\rho\sigma}R^{\mu\nu\rho\sigma}=\frac{m^2 \left(36 a^4-48 a^2 r^2+47 r^4\right)+32 m a^2 \left(r^2-a^2\right) \sqrt{a^2+r^2}+12 a^4 \left(a^2+r^2\right)}{\left(a^2+r^2\right)^5}.
\end{equation}
The above equation clearly demonstrates that the Kretschmann scalar maintains its finite value at $r=0$ unless $a$ is equal to zero. Consequently, the magnetic charge $a$ serves as a regulating factor that prevents the occurrence of a singularity at the center of the black hole. This indicates that the solution represents a black bounce, which is a black hole without any singularities and also possesses a wormhole solution.

The black bounce model proposed by Simpson-Visser fails to satisfy Einstein's equations in a vacuum when considering GR. Specifically, the matter content can be interpreted as an anisotropic fluid.
 \begin{eqnarray}
 	\rho&=&-\frac{a^{2}(\sqrt{r^2+a^{2}}-4m)}{8\pi(r^{2}+a^{2})^{5/2}},\\
 	p_{1}&=&-\frac{a^{2}}{8\pi(r^{2}+a^{2})^{2}},\\
 	p_{2}&=&\frac{a^{2}(\sqrt{r^2+a^{2}}-m)}{8\pi(r^{2}+a^{2})^{5/2}},
 \end{eqnarray}
The stress energy tensor is
\begin{equation}
	T^{\mu}_{\nu}=diag[-\rho, p_{1},p_{2},p_{2}],
\end{equation}
where we used the $t$ as timelike.

To account for the geodesic memory effects in the Simpson-Visser background when a GW pulse is present, we consider the metric as the combination of the background metric $g_{\mu\nu}$, given by equation $(\ref{bouncemetric})$, and the perturbation caused by the GW pulse, denoted as $h_{\mu\nu}$. However, for convenience, we first apply the transformation $R^{2}=r^{2}+a^{2}$ to the metric $(\ref{bouncemetric})$. This leads us to the following expression for the line element:
\begin{equation}\label{metricR}
    ds^{2}=-F(R)dt^{2}+G(R)dR^{2}+R^{2}(d\theta^{2}+\sin^{2}\theta d\phi^{2}),
\end{equation}
where 
\begin{equation}
    F(R)=1-\frac{2m}{R},~~~~~~~~~~ G(R)= \frac{R^{3}}{(R-2m)(R^{2}-a^{2})}.
\end{equation}
In the following we use the transformation $u=t-r_{*}$ where $r_{*}$ is the tortoise coordinate and is defined as 
	\begin{align}\label{tor}
   &~~~~~~~~~~~~~~~~~~~~~ \frac{d r_{*}}{dR}=\sqrt{\frac{G(R)}{F(R)}},\\
\label{metricRR}
    ds^{2}&=-F(R)du^{2}-2\sqrt{G(R)F(R)}dudR+R^{2}(d\theta^{2}+\sin^{2}\theta d\phi^{2}).
\end{align}

For our convenience we use the expression  $Z(R)=\sqrt{G(R)F(R)}$.  Perturbation of the metric $(\ref{metricRR})$ leads to
\begin{equation}\label{perturb}
    ds^{2}=(g_{\mu\nu}+h_{\mu\nu})dx^{\mu}dx^{\nu},
\end{equation}
Then the resulting geometry becomes
\begin{equation}\label{metricRRR}
    ds^{2}=-F(R)du^{2}-2Z(R)dudR +(R^{2}+RH(u))d\theta^{2}+(R^{2}-RH(u))\sin^{2}\theta d\phi^{2}.
\end{equation}
We define $H(u)$ as the GW pulse, which has the form
\begin{equation}\label{Hu}
    H(u)=A~\text{sech} ^{2}(u-u_{0}), 
\end{equation}
where $A$ is the amplitude of the GW pulse and $u_{0}$ is its center. On the equatorial plane $(\theta=\frac{\pi}{2})$, the geodesic equation for the $u$ coordinate is
\begin{equation}\label{14}
    \ddot{u}-\frac{F^{\prime}}{2Z}\dot{u}^{2}-\frac{H-2R}{2Z}\dot{\phi}^{2}=0,
\end{equation}
where $F$, $Z$, and $R$ are functions of $u$ and $\phi$. The geodesic equations for $r$ and $\phi$ are
\begin{align}
	\label{15}
    &~~~~~~~~~~~~~~~~~\ddot{R}+\frac{F}{2Z^{2}}F^{\prime}\dot{u}^{2}+\frac{F^{\prime}}{Z}\dot{R}\dot{u}+\frac{Z^{\prime}}{Z}\dot{R}^{2}+\frac{FH-2FR-RZH^{\prime}}{2Z^{2}}\dot{\phi}^{2}=0,&\\
    \label{16}
   &~~~~~~~~~~~~~~~~~~~~~~~~~~~~~~~~~\ddot{\phi}-\frac{H^{\prime}}{R-H} \dot{\phi}\dot{u}+\frac{2R-H}{R(R-H)}\dot{\phi}\dot{R}=0,&
\end{align}
    where $H^{\prime}=\frac{dH}{du}$ and $F^{\prime}=\frac{dF}{dR}$. 

The background geometry which is the black bounce or , in the next section, can be a traversable wormhole solution, considering for large $r$, on the equatorial plane $\theta=\frac{\pi}{2}$ satisfies the following condition
\begin{equation}\label{17}
    -F(R) \dot{u}^{2}-2Z(R)\dot{u}\dot{R}+R^{2}\dot{\phi}^{2}-RH(u)\dot{\phi}^{2}=-1.
\end{equation}
In order to determine two neighboring geodesics in the presence of a GW pulse, it is necessary to employ numerical methods to solve the non-linear equations $(\ref{14})$, $(\ref{15})$, and $(\ref{16})$. The initial conditions for these geodesics are chosen as follows: the initial values of $r$ and $u$ for geodesics I and II can be arbitrary, but the value of $\phi$ remains the same for both. Furthermore, the initial conditions for $r$ and $u$ are fixed for the geodesics. Since we are confined to the equatorial plane, the value of $\phi$˙ can be determined by utilizing equation $(\ref{17})$.

The equations provided express the calculation of the disparity between two geodesics. This disparity is determined by subtracting the values of "$r$" and "$u$" for the second geodesic from the corresponding values for the first geodesic.
\begin{equation}
  \Delta R=R(geodesic II)-R(geodegic I) ,~~~~~~ \Delta u=u(geodesic II)-u(geodesic I).  
\end{equation}

%\begin{figure}[!tbp]
\begin{figure}[H]
	\centering
	\begin{minipage}[b]{0.4\textwidth}
		\includegraphics[width=\textwidth]{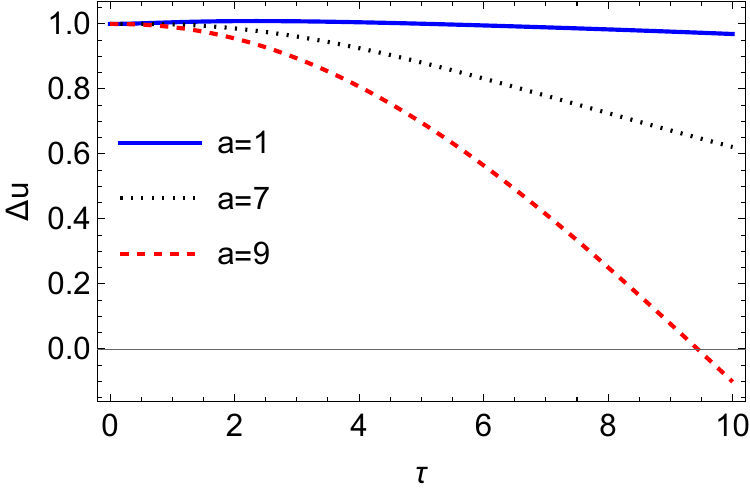}
		\caption{ Variation of $\Delta u$ due to GW pulse has been presented with respect to   $\tau$, for different choices of $a$ and certain values of $m=1$ and $A_{0}=1$ for Simpson-Visser black bounce.}
		\label{fig1}
	\end{minipage}
	\hfill
	\begin{minipage}[b]{0.4\textwidth}
		\includegraphics[width=\textwidth]{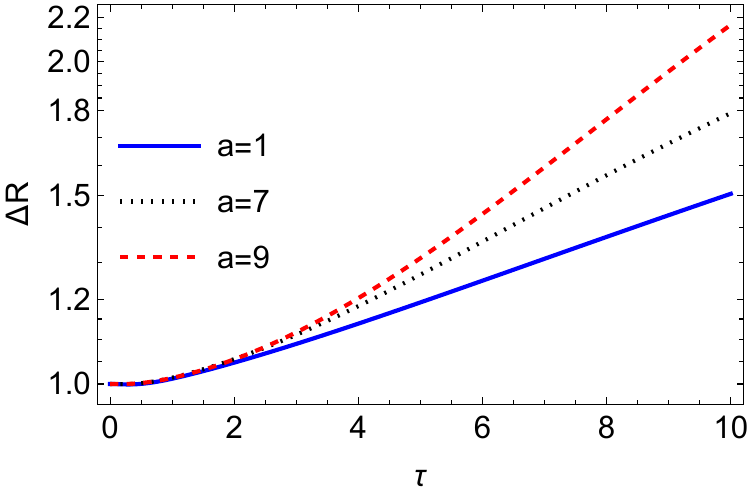}
		\caption{ Variation of $\Delta R$ due to GW pulse with respect to $\tau$, for different choices of $a$ and certain values of $m=1$ and $A_{0}=1$ for Simpson-Visser black bounce.}
		\label{fig2}
	\end{minipage}
\end{figure}

Figures \ref{fig1} and \ref{fig2} portray the values of $\Delta u$ and $\Delta R$, for different values of magnetic charge $a$ and and certain values of $m=1$ and $A_{0}=1$, correspondingly, for the Simpson-Visser solution represented as a black bounce. The figures demonstrate that as the magnetic charge $a$ increases, the difference between two adjacent geodesics for the component $u$ ($\Delta u$) diminishes. Conversely, for the component $R$, the difference increases with an augmentation in the magnetic charge $a$, as exhibited in figure \ref{fig2}.

  \begin{figure}[H]
	\begin{minipage}[b]{0.4\textwidth}
		\includegraphics[width=\textwidth]{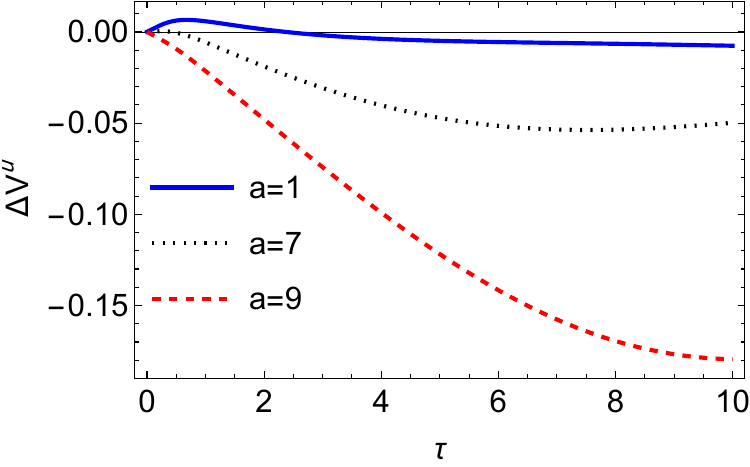}
		\caption{ Variation of $\Delta V^{u}$ due to GW pulse with respect to $\tau$,  for different choices of $a$ and certain values of $m=1$ and $A_{0}=1$, for Simpson-Visser black bounce.}
		\label{fig3}
	\end{minipage}
	\hfill
	\begin{minipage}[b]{0.4\textwidth}
		\includegraphics[width=\textwidth]{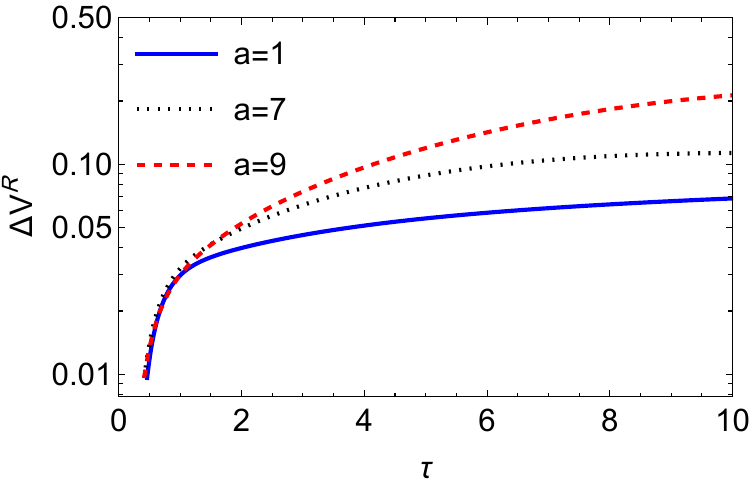}
		\caption{ Variation of $\Delta V^{R}$ due to GW pulse with respect to $\tau$, for different choices of $a$ and certain values of $m=1$ and $A_{0}=1$, for Simpson-Visser black bounce.}
		\label{fig4}
	\end{minipage}
\end{figure}

These disparities in $R$ and $u$ can be interpreted as a displacement memory effect. The velocity memory effect refers to the changes in velocities of $u$ and $R$, which are determined by the variations of $\Delta V^{u}$ and $\Delta V^{R}$ with respect to $\tau$ and for certain values of $m=1$ and $A_{0}=1$, as illustrated in figures \ref{fig3} and \ref{fig4}, respectively.

%\begin{figure}[!tbp]
\begin{figure}[H]
	\centering
	\begin{minipage}[b]{0.4\textwidth}
		\includegraphics[width=\textwidth]{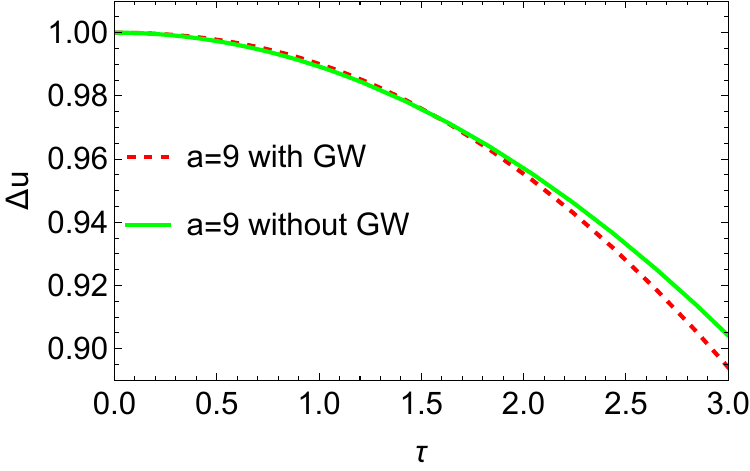}
		\caption{Variation of $\Delta u$ due to GW pulse and without GW pulse with respect to $\tau$ for certain values of $m=1$ and $A_{0}=1$ of Simpson-Visser black bounce.}
		\label{fig5}
	\end{minipage}
	\hfill
	\begin{minipage}[b]{0.4\textwidth}
		\includegraphics[width=\textwidth]{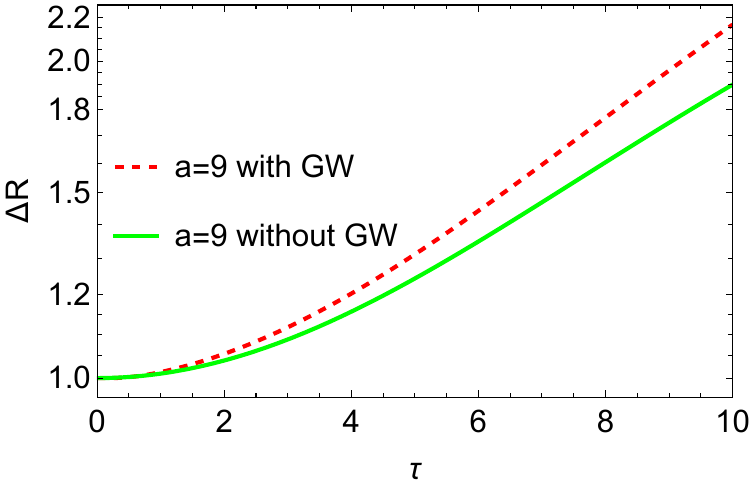}
		\caption{ Variation of $\Delta R$ due to GW pulse and without GW pulse with respect to $\tau$ for certain values of $m=1$ and $A_{0}=1$ of Simpson-Visser black bounce.}
		\label{fig6}
	\end{minipage}
\end{figure}

\begin{figure}[H]
	\begin{minipage}[b]{0.4\textwidth}
		\includegraphics[width=\textwidth]{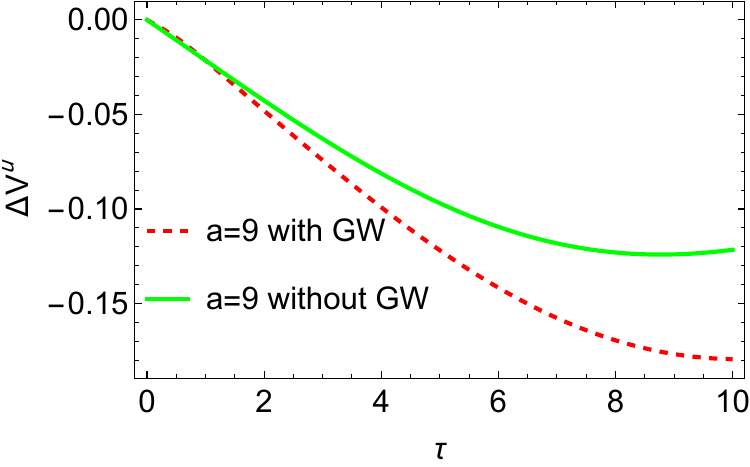}
		\caption{Variation of $\Delta V^{u}$ due to GW pulse and without GW pulse with respect to $\tau$ for certain values of $m=1$ and $A_{0}=1$ of Simpson-Visser black bounce.}
		\label{fig7}
	\end{minipage}
	\hfill
	\begin{minipage}[b]{0.4\textwidth}
		\includegraphics[width=\textwidth]{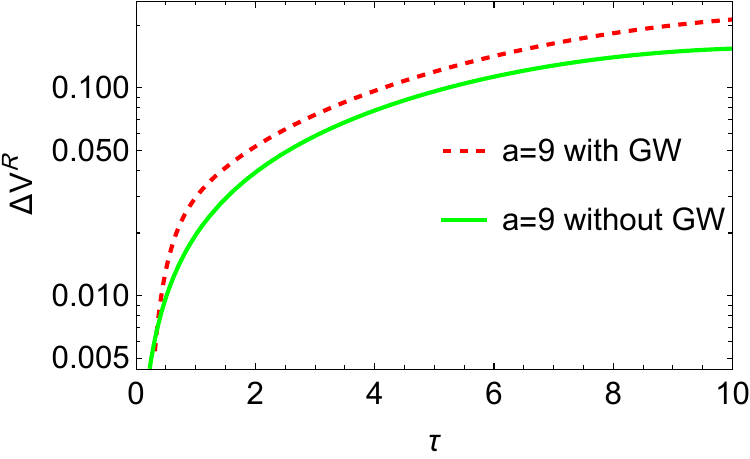}
		\caption{ Variation of $\Delta V^{R}$ due to GW pulse and without GW pulse with respect to $\tau$ for certain values of $m=1$ and $A_{0}=1$ of Simpson-Visser black bounce.}
		\label{fig8}
	\end{minipage}
\end{figure}

Figures \ref{fig5} to \ref{fig8} provide a clear demonstration of the tangible GW pulse in the context of the Simpson-Visser black bounce. These figures showcase both the displacement and velocity memory effects. Specifically, figures \ref{fig5} and \ref{fig6} depict the changes in displacement memory effects when the GW pulse is present or absent for magnetic charge $a=9$. On the other hand, figures \ref{fig7} and \ref{fig8} illustrate the variations in $\Delta V^{u}$ and $\Delta V^{R}$ in the presence or absence of the GW pulse for certain values of $m=1$ and $A_{0}=1$, again for magnetic charge $a=9$.

\section{Geodesic memory effect for Bardeen-Type solution as a black bounce}
In this section we consider the gravitational memory effects in the Bardeen-type black bounce background. 
The Bardeen-type black bounce solution described by the line element $(\ref{bouncemetric})$ and with
\begin{equation}
     f(r)=1-\frac{2mr^{2}}{(r^{2}+a^{2})^{3/2}}, ~~~and~~~ \sigma (r)=\sqrt{r^{2}+a^{2}}.
\end{equation}
In our previous discussion on the Simpson-Visser black bounce, we observed that the parameter $a$, representing the magnetic charge, plays a crucial role in preventing the emergence of a central singularity at $r=0$. Likewise, in the case of the Bardeen-Type solution, we encounter a parameter $a$ that serves as a regularization factor, effectively averting the formation of a singularity at $r=0$. The Kretschmann scalar associated with this solution can be expressed as follows.
\begin{eqnarray}
 K=\frac{m^2 \left(16 a^8-176 a^6 r^2+672 a^4 r^4-268 a^2 r^6+47 r^8\right)-32 m a^2 r^2 \sqrt{a^2+r^2} \left(2 a^4+a^2 r^2-r^4\right)+12 a^4 \left(a^2+r^2\right)^3}{\left(a^2+r^2\right)^7}.   
\end{eqnarray}
The finiteness of the Kretschmann scalar at $r=0$ is evident from the aforementioned expression, provided that $a \neq 0$. Hence, the parameter $a$ assumes a comparable significance to that in the Simpson-Visser scenario, thereby characterizing the solution as a black bounce.

In the Simpson-Visser solution, the discussion highlighted the need to consider geodesic memory effects in the Bardeen-Type background when a GW pulse is present. To account for this, the metric is expressed as a combination of the background metric, denoted as $g_{\mu\nu}$ and described by equation $(\ref{bouncemetric})$, and the perturbation caused by the GW pulse, represented as $h_{\mu\nu}$. To simplify the analysis, the transformation $R^{2}=r^{2}+a^{2}$ is applied to the metric $(\ref{bouncemetric})$. This transformation leads to the line element $(\ref{metricR})$ with the corresponding functions.
\begin{equation}\label{F}
    F(a,R)=1-\frac{2m}{R}+\frac{2ma^{2}}{R^{3}}, 
\end{equation}
and 
\begin{equation}\label{G}
    G(a,R)= \frac{R^{5}}{(R^{3}-2mR^{2}+2ma^{2})(R^{2}-a^{2})}.
\end{equation}
we define $Z(a,R)=\sqrt{G(a,R)F(a,R)}$ as follows
\begin{equation}\label{Z}
    Z(a,R)=\sqrt{\frac{R^{2}}{R^{2}-a^{2}}}.
\end{equation}
In this case, the metric $(\ref{metricRR})$ incorporates the variable $Z$. The definition of this metric relies on the transformation $u=t-r_{*}$, where $r_{*}$ represents the tortoise coordinate. The tortoise coordinate is determined by the transformation $(\ref{tor})$.

In this particular scenario, the perturbation of the metric $(\ref{metricRR})$ also results in a modified line element $(\ref{metricRRR})$. Additionally, within this model, we have defined $H(u)$ as the GW pulse, which can be expressed as $(\ref{Hu})$. On the equatorial plane $(\theta=\pi/2)$, the geodesic equation for the $u$, $R$, and $\phi$ coordinates are given by equations $(\ref{14})$, $(\ref{15})$, and $(\ref{16})$, respectively. However, it is important to note that these equations incorporate the definitions $(\ref{F})$, $(\ref{G})$, and $(\ref{Z})$ for the functions $F$, $G$, and $Z$, respectively.

%\begin{figure}[!tbp]
\begin{figure}[H]
	\centering
	\begin{minipage}[b]{0.4\textwidth}
		\includegraphics[width=\textwidth]{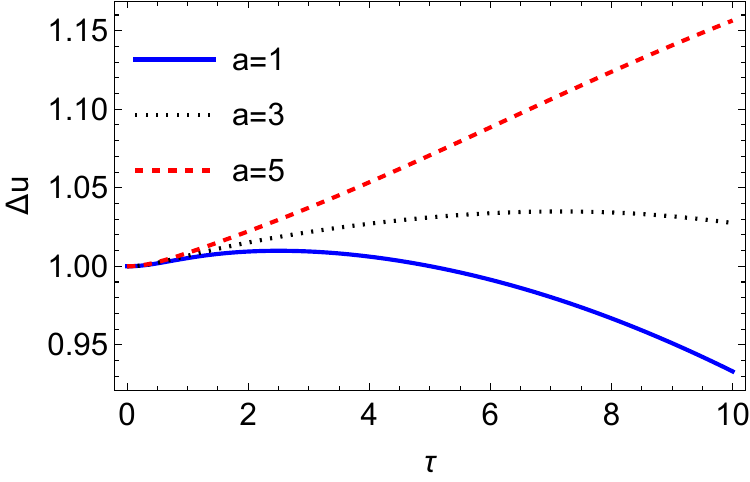}
		\caption{ Variation of $\Delta u$ due to GW pulse has been presented with respect to   $\tau$, for different choices of $a$ and certain values of $m=1$ and $A_{0}=1$ for Bardeen-Type black bounce.}
		\label{fig9}
	\end{minipage}
	\hfill
	\begin{minipage}[b]{0.4\textwidth}
		\includegraphics[width=\textwidth]{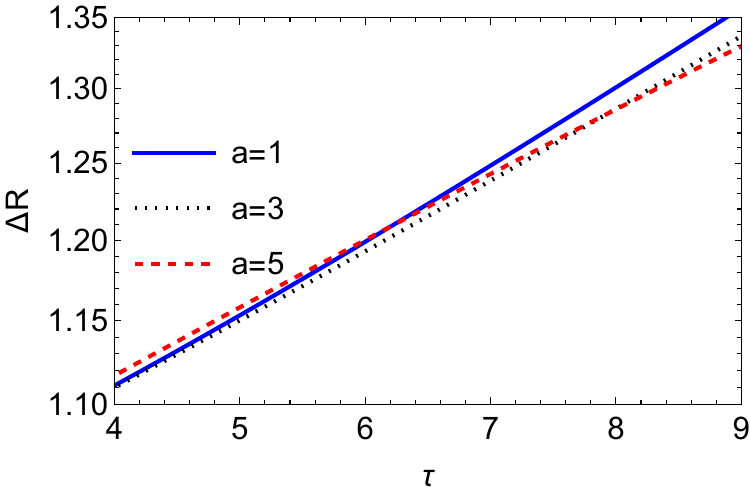}
		\caption{ Variation of $\Delta R$ due to GW pulse with respect to $\tau$, for different choices of $a$ and certain values of $m=1$ and $A_{0}=1$ for Bardeen-Type black bounce.}
		\label{fig10}
	\end{minipage}
\end{figure}

In the context of the Simpson-Visser solution, the displacement and velocity memory effects are also applicable to the Bardeen-Type black bounce. The Bardeen-Type solution, depicted as a black bounce, exhibits similar characteristics. Specifically, Figures \ref{fig9} and $(10)$ illustrate the values of $\Delta u$ and $\Delta R$ for certain values of $m=1$ and $A_{0}=1$, respectively. These figures demonstrate that as the magnetic charge $a$ increases, the difference between two adjacent geodesics for the component $u$ the displacement ($\Delta u$) also increases. Conversely, for the component $R$, the difference decreases as the magnetic charge $a$ increases, as shown in figure \ref{fig10}.

  \begin{figure}[H]
	\begin{minipage}[b]{0.4\textwidth}
		\includegraphics[width=\textwidth]{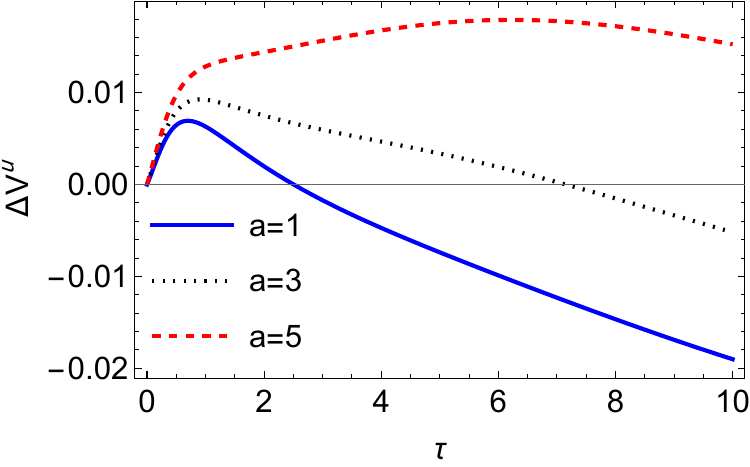}
		\caption{ Variation of $\Delta V^{u}$ due to GW pulse with respect to $\tau$, for different choices of $a$ and certain values of $m=1$ and $A_{0}=1$ for Bardeen-Type black bounce.}
		\label{fig11}
	\end{minipage}
	\hfill
	\begin{minipage}[b]{0.4\textwidth}
		\includegraphics[width=\textwidth]{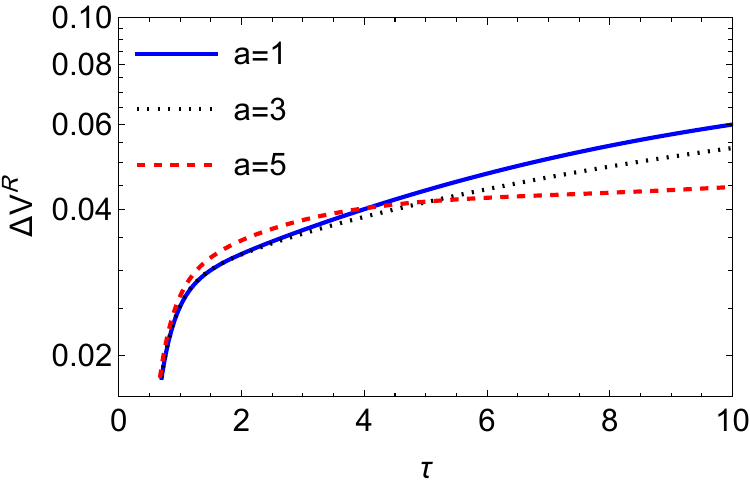}
		\caption{ Variation of $\Delta V^{R}$ due to GW pulse with respect to $\tau$, for different choices of $a$ and certain values of $m=1$ and $A_{0}=1$ for Bardeen-Type black bounce.}
		\label{fig12}
	\end{minipage}
\end{figure}

The changes observed in $R$ and $u$ can be interpreted as a memory effect related to displacement. On the other hand, the memory effect associated with velocity refers to the variation in the velocities of $u$ and $R$. These differences are determined by the variations of $\Delta V^{u}$ and $\Delta V^{R}$ with respect to $\tau$, as depicted in figures \ref{fig11} and \ref{fig12} for different values of magnetic charge $a$ and for certain values of $m=1$ and $A_{0}=1$, respectively. It is worth noting that the changes in magnetic charge $a$ have a similar impact on the velocity memory effects as they do on the displacement memory effects.

%\begin{figure}[!tbp]
\begin{figure}[H]
	\centering
	\begin{minipage}[b]{0.4\textwidth}
		\includegraphics[width=\textwidth]{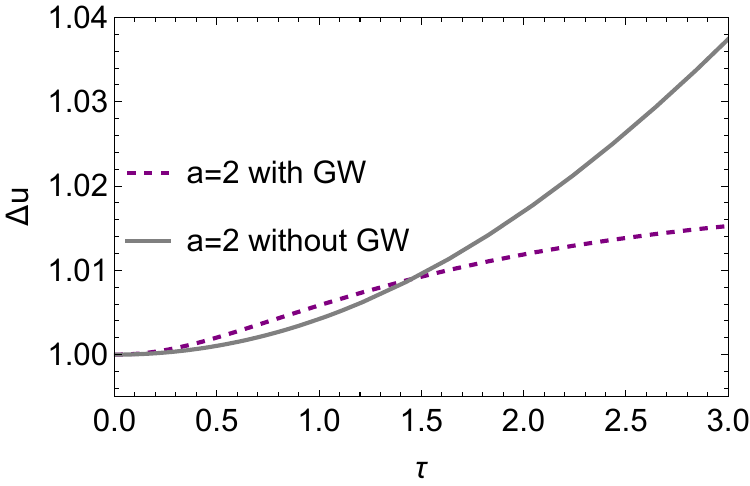}
		\caption{Variation of $\Delta u$ due to GW pulse and without GW pulse with respect to $\tau$ and for certain values of $m=1$ and $A_{0}=1$ for Bardeen-Type black bounce.}
		\label{fig13}
	\end{minipage}
	\hfill
	\begin{minipage}[b]{0.4\textwidth}
		\includegraphics[width=\textwidth]{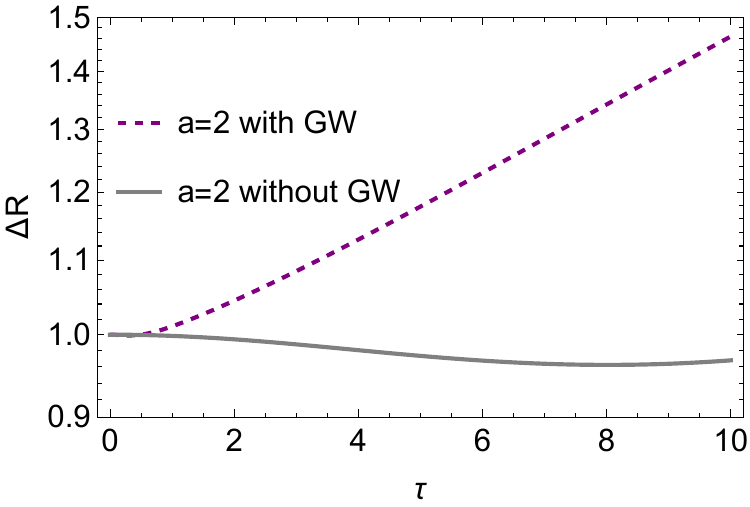}
		\caption{ Variation of $\Delta R$ due to GW pulse and without GW pulse with respect to $\tau$ and for certain values of $m=1$ and $A_{0}=1$ for Bardeen-Type black bounce.}
		\label{fig14}
	\end{minipage}
	
\end{figure}
\begin{figure}[H]
	\begin{minipage}[b]{0.4\textwidth}
		\includegraphics[width=\textwidth]{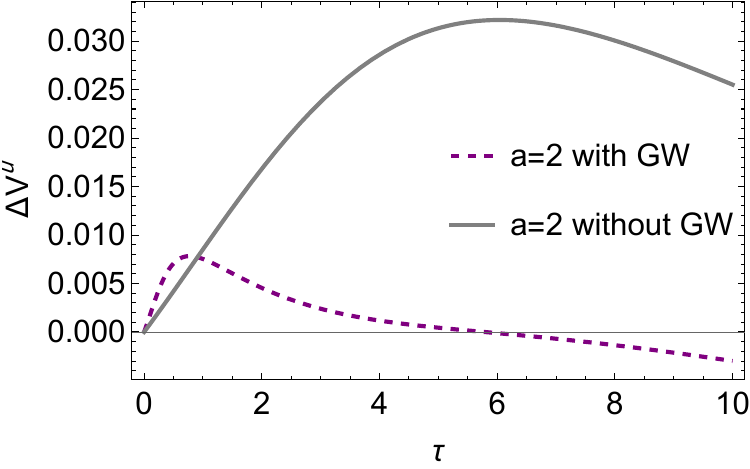}
		\caption{Variation of $\Delta V^{u}$ due to GW pulse and without GW pulse with respect to $\tau$ and for certain values of $m=1$ and $A_{0}=1$ for Bardeen-Type black bounce.}
		\label{fig15}
	\end{minipage}
	\hfill
	\begin{minipage}[b]{0.4\textwidth}
		\includegraphics[width=\textwidth]{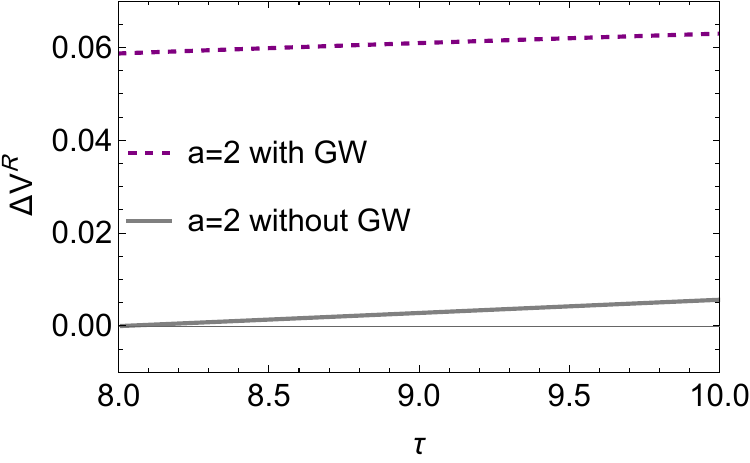}
		\caption{ Variation of $\Delta V^{R}$ due to GW pulse and without GW pulse with respect to $\tau$ and for certain values of $m=1$ and $A_{0}=1$ for Bardeen-Type black bounce.}
		\label{fig16}
	\end{minipage}
\end{figure}

Figures \ref{fig13} through \ref{fig16} offer a clear demonstration of the tangible GW pulse within the context of the Bardeen-Type black bounce. These figures showcase both the displacement and velocity memory effects. Specifically, figures \ref{fig13} and \ref{fig14} portray the changes in displacement memory effects when the GW pulse is either present or absent for a magnetic charge of $a=2$ and certain values of $m=1$ and $A_{0}=1$. Conversely, figures \ref{fig15} and \ref{fig16} illustrate the variations in $\Delta V^{u}$ and $\Delta V^{R}$ in the presence or absence of the GW pulse, once again for a magnetic charge of $a=2$. The geodesic deviation of two neighboring points decreases for the $u$ component of displacement and velocity memory effects when the GW pulse is present. However, this physical phenomenon is vice versa for the $R$ component.

\section{A traversable wormhole solution and memory effects}
Given that black bounce models already have wormhole solutions, our focus is now on exploring other wormhole solutions to examine their memory effects. In this particular section, we will investigate the gravitational memory effects of a special type of wormhole solution known as a traversable wormhole. To accomplish this, we will analyze the Einstein-Dirac-Maxwell (EDM) model. This model involves two gauged relativistic fermions, with opposite spins to maintain spherical symmetry. For a more comprehensive understanding of this model, readers can refer to the source \cite{26}. In our analysis, we will work with units where $G=c=\hslash =1$, and the action of the corresponding EDM model can be expressed accordingly.
\begin{equation}\label{action0}
    S=\frac{1}{4\pi}\int d^{4}x \sqrt{-g}\left[\frac{R}{4}+ L_{D}-\frac{F^{2}}{4}\right],
\end{equation}
The Dirac Lagrangian density for a charged spinor field coupled to an electromagnetic field is given by:
\begin{equation}\label{dirac}
L_{D}=\sum_{\epsilon =1,2}\left[\frac{i}{2}\bar{\psi}{\epsilon}\gamma^{\nu}D{\nu}\psi_{\epsilon}-\frac{i}{2}D_{\nu}\bar{\psi}{\epsilon}\gamma^{\nu}\psi{\epsilon}-m\bar{\psi}{\epsilon}\psi{\epsilon}\right],
\end{equation}

 $R$ is the Ricci scalar of the metric $g_{\mu\nu}$, $F_{\mu\nu}=\partial_{\mu}A_{\nu}-\partial_{\nu}A_{\mu}$ is the field strength tensor, $L_{D}$ is the Dirac Lagrangian density, $\psi_{\epsilon}$ is the spinor field, $\gamma^{\nu}$ is the curved space $\gamma$ matrix,  $D_{\mu}=\partial_{\mu}+\Gamma_{\mu}-ieA_{\mu}$ is the covariant derivative where $\Gamma_{\mu}$ and $e$ are the spinor connection matrices and the gauge coupling constant, respectively. $m$ is the mass of both spinors, and $i$ is the imaginary unit. Equation $(\ref{dirac})$ describes the interaction between the spinor field and the electromagnetic field in a curved spacetime.

By variation of action $(\ref{action0})$  with respect to the metric and fermion fields, the field equations are given by
\begin{align}
    R_{\mu\nu}-\frac{1}{2}Rg_{\mu\nu}=2T_{\mu\nu}, ~~~~~~~ T_{\mu\nu}=T_{\mu\nu}^{D}+T_{\mu\nu}^{M},\\
    (\gamma^{\nu}D_{\nu}-m)\psi_{\epsilon}=0,   ~~~~~~and ~~~~~\nabla_{\mu}F^{\mu\nu}=ej^{\nu}.
\end{align}
In this context, the symbol $j^{\nu}$ represents the current, which is obtained by summing over the values of $\epsilon$ (which can be either 1 or 2) and multiplying the corresponding fermion fields $\bar{\psi}_{\epsilon}$, $\gamma^{\nu}$, and $\psi_{\epsilon}$
\begin{equation}
    j^{\nu}=\sum_{\epsilon =1,2}\bar{\psi}_{\epsilon}\gamma^{\nu}\psi_{\epsilon}.
\end{equation}
On the other hand, the energy tensor $T_{\mu\nu}^{D}$ for the fermion field is calculated by summing over the values of $\epsilon$ (again, either 1 or 2), multiplying the imaginary part of $\bar{\psi}_{\epsilon}$, $\gamma_{(\mu}D_{\nu)}$, and $\psi_{\epsilon}$ by 2.
\begin{equation}
    T_{\mu\nu}^{D}=\sum_{\epsilon= 1,2}2Im(\bar{\psi}_{\epsilon}\gamma_{(\mu}D_{\nu)}\psi_{\epsilon}).
\end{equation}
Furthermore, the energy tensor for the Maxwell part is denoted as $T_{\mu\nu}^{M}$ and is determined by the expression $F_{\mu\alpha}F^{\alpha}_{\nu}-\frac{1}{4}F^{2}g_{\mu\nu}$, where $F_{\mu\alpha}$ represents the electromagnetic field tensor and $g_{\mu\nu}$ is the metric tensor.

Considering only static, spherically symmetric solutions of the field equations, we examine a comprehensive metric. We assume that the field $A$ is solely electric, given by $A=V(r)dt$, where $r$ and $t$ represent the radial and time coordinates, respectively. By imposing the conditions $e=0$, with vanishing frequency $\omega=0$, and assuming the massless spinor fields, the resulting EDM equations can be analytically solved. The solution encompasses the metric and the $U(1)$ potential \cite{26} in the following manner.
\begin{align}\label{metricWH}
    &ds^{2}=-(1-\frac{\mathcal{M}}{r})^{2}dt^{2}+\frac{dr^{2}}{(1-\frac{r_{0}}{r})(1-\frac{Q_{e}^{2}}{r_{0}r})}+r^{2}d\Omega^{2},\\
    &~~~~~~~~~~V(r)=\frac{\mathcal{M}}{Q_{e}}\sqrt{(1-\frac{r_{0}}{r})(1-\frac{Q_{e}^{2}}{r_{0}r})},
\end{align}
where $\mathcal{M}$ is defined as
\begin{equation}
    \mathcal{M}=\frac{2Q_{e}^{2}r_{0}}{Q_{e}^{2}+r_{0}^{2}}.
\end{equation}
This passage discusses a traversable wormhole (WH) solution, where the throat's radius is denoted as $r_{0}$ and the electric charge $Q_{e}$ is less than $r_{0}$.
with eleetrie charge. 
 The WH solution is characterized by the Arnowitt-Deser-Misner (ADM) mass $M$, with the condition that $Q_{e}/\mathcal{M} > 1$. The WH geometry is sustained by the contribution of spinors to the total energy-momentum tensor, ensuring regularity throughout. As $Q_{e}$ approaches $r_{0}$, the WH solution transitions towards the extremal Reissner-Nordström (RN) black hole, while the contribution of $T_{\mu\nu}^{D}$ tends towards zero.

In this specific traversable Wormhole scenario, the conversation emphasized the importance of taking into account geodesic memory effects as displacement and velocity memory effects in the black bounce background (Simpson-Visser and Bardeen-Type) when a GW pulse is present. To address this, the metric is formulated as a combination of the background metric, denoted as $g_{\mu\nu}$ and defined by equation $(\ref{metricWH})$, and the perturbation induced by the GW pulse, represented as $h_{\mu\nu}$.

In order to achieve our objective, we proceed to the tortoise coordinate as in previous cases, which leads to the following line element:
\begin{equation}\label{tortoiseWH}
    ds^{2}=-f(r)du^{2}-2Z(r)dudr+r^{2}d\theta^{2}+r^{2}\sin^{2}\theta d\phi^{2},
\end{equation}
Here, $Z(r)=\sqrt{g(r)f(r)}$ and the functions $f(r)$ and $g(r)$ are defined as:
\begin{equation}
    f(r)=(1-\frac{\mathcal{M}}{r})^{2}, ~~~and~~~ g(r)=\frac{1}{(1-\frac{r_{0}}{r})(1-\frac{Q_{e}^{2}}{r_{0}r})}.
\end{equation}
If we introduce the perturbation metric $h_{\mu\nu}$ to the background metric $(\ref{tortoiseWH})$, we obtain:
\begin{equation}
    ds^{2}=-f(r)du^{2}-2Z(r)dudr+(r^{2}+rH(u))d\theta^{2}+(r^{2}-rH(u))\sin^{2}\theta d\phi^{2}.
\end{equation}
In this case as well, we define $H(u)$ as the GW pulse, which takes the form $(\ref{Hu})$. 

On the equatorial plane $(\theta=\frac{\pi}{2})$, the geodesic equations for the $u$ coordinate become: 
\begin{equation}\label{WHu}
    \ddot{u}-\frac{f^{\prime}}{2Z}\dot{u}^{2}-\frac{H-2r}{2Z}\dot{\phi}^{2}=0,
\end{equation}
Here, $f$, $Z$, and $r$ are functions of $u$ and $\phi$. The geodesic equations for $r$ and $\phi$ are given by:
\begin{align}\label{WHr}
    &\ddot{r}+\frac{f}{2Z^{2}}f^{\prime}\dot{u}^{2}+\frac{f^{\prime}}{Z}\dot{r}\dot{u}+\frac{Z^{\prime}}{Z}\dot{r}^{2}+\frac{fH-2fr-rZH^{\prime}}{2Z^{2}}\dot{\phi}^{2}=0,\\
&\label{WHphi}
   ~~~~~~~~~~~~~~~~~\ddot{\phi}-\frac{H^{\prime}}{r-H} \dot{\phi}\dot{u}+\frac{2r-H}{r(r-H)}\dot{\phi}\dot{r}=0,
\end{align}
    where $H^{\prime}=\frac{dH}{du}$ and $f^{\prime}=\frac{df}{dr}$. 
   
 In our investigation of displacement and velocity memory effects, similar to previous cases, we focus on the deviation between two neighboring geodesics. In this particular study, we aim to explore the influence of electric charge, denoted as $Q_{e}$, on the gravitational memory effect. The reason behind considering the electric charge and its impact on gravitational memory lies in the unique nature of this charge. It is important to note that there is no presence of any "real" charge in the given scenario. Instead, we have a purely geometric representation of electric charge, which is described in terms of the metric and topology of curved empty space. In other words, it is a charge without any physical charge. Therefore, the detection of traces of this type of charge in the gravitational memory serves as evidence supporting Wheeler's concept of "charge without charge" \cite{wheeler}.

 In this particular instance of traversable wormhole solution, similar to our previous solutions for black bounces, we observe the displacement memory effect denoted as $\Delta r$ and $\Delta u$, as well as the velocity memory effects represented by $\Delta V^{r}$ and $\Delta V^{u}$. These effects are obtained through numerical solutions of the equations $(\ref{WHu})$, $(\ref{WHr})$, and $(\ref{WHphi})$. Consequently, within this framework, we obtain the subsequent outcomes.
 
 %\begin{figure}[!tbp]
 \begin{figure}[H]
 	\centering
 	\begin{minipage}[b]{0.4\textwidth}
 		\includegraphics[width=\textwidth]{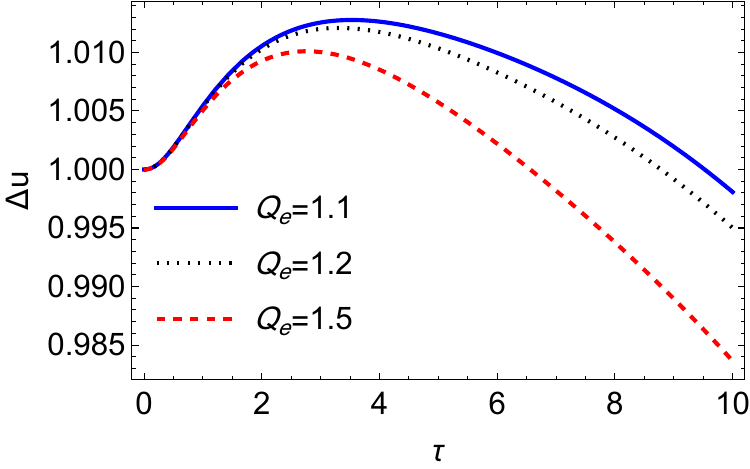}
 		\caption{ The change in $\Delta u$ resulting from a GW pulse in relation to $\tau$ was demonstrated, taking into account different values of $Q_{e}$ of TWH. The parameters were specified as  $\mathcal{M}=1$, $A_{0}=1$ and $r_{0}=2$ for the figure presentation.}
 		\label{fig17}
 	\end{minipage}
 	\hfill
 	\begin{minipage}[b]{0.4\textwidth}
 		\includegraphics[width=\textwidth]{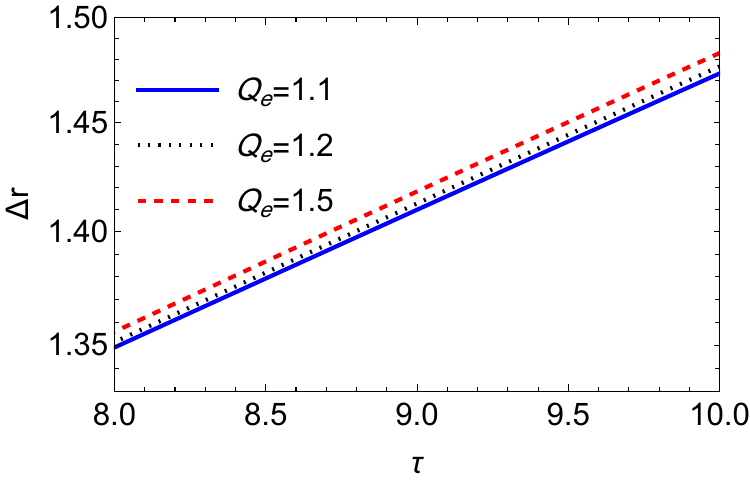}
 		\caption{ The change in $\Delta r$ resulting from a GW pulse in relation to $\tau$ was demonstrated, taking into account different values of $Q_{e}$ of TWH. The parameters were specified as $\mathcal{M}=1$, $A_{0}=1$ and $r_{0}=2$ for the figure presentation.}
 		\label{fig18}
 	\end{minipage}
 \end{figure}

 Figures \ref{fig17} and \ref{fig18} depict the variations in $\Delta u$ and $\Delta r$ for different values of $Q_{e}$ of  traversable wormhole (TWH) for the figure presentation correspondingly . The parameters were specified as $\mathcal{M}=1$, $A_{0}=1$ and $r_{0}=2$. These visual representations provide evidence that the disparity between two neighboring geodesics for the displacement component $u$ ($\Delta u$) diminishes with an increase in the electric charge $Q_{e}$. Conversely, for the component $r$, the difference decreases as the electric charge $Q_{e}$ decreases, as illustrated in figure \ref{fig18}.

  \begin{figure}[H]
 	\begin{minipage}[b]{0.4\textwidth}
 		\includegraphics[width=\textwidth]{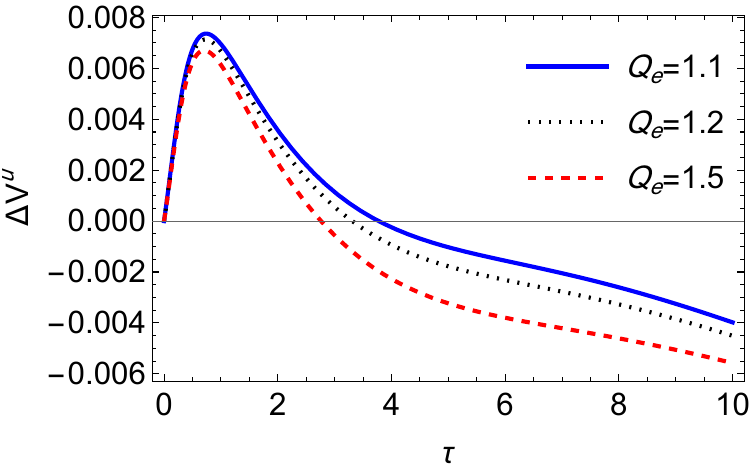}
 		\caption{ Variation of $\Delta V^{u}$ due to GW pulse with respect to $\tau$, for different choices of $Q_{e}$ of TWH.( $\mathcal{M}=1$, $A_{0}=1$ and $r_{0}=2$)}
 		\label{fig19}
 	\end{minipage}
 	\hfill
 	\begin{minipage}[b]{0.4\textwidth}
 		\includegraphics[width=\textwidth]{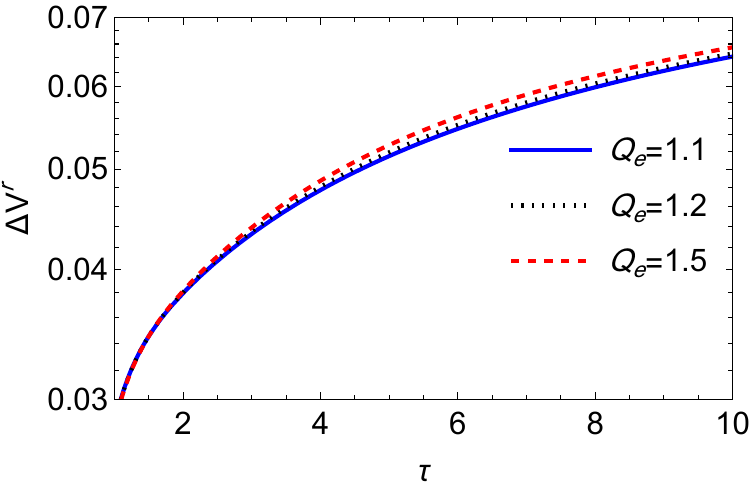}
 		\caption{ Variation of $\Delta V^{r}$ due to GW pulse with respect to $\tau$, for different choices of $Q_{e}$ of TWH.( $\mathcal{M}=1$,$A_{0}=1$ and $r_{0}=2$)}
 		\label{fig20}
 	\end{minipage}
 \end{figure}

The changes witnessed in $r$ and $u$ can be explained as a memory phenomenon linked to displacement. Conversely, the memory effect connected to velocity pertains to the variations in the velocities of $u$ and $r$. These disparities are influenced by the changes in $\Delta V^{u}$ and $\Delta V^{r}$ in relation to $\tau$, as illustrated in figures \ref{fig19} and \ref{fig20} for different values of $Q_{e}$ and certain parameters  $\mathcal{M}=1$, $A_{0}=1$ and $r_{0}=2$, respectively. It is important to highlight that the modifications in electric charge $Q_{e}$ have a comparable influence on the memory effects of velocity as they do on the memory effects of displacement. 

%\begin{figure}[!tbp]
\begin{figure}[H]
	\centering
	\begin{minipage}[b]{0.4\textwidth}
		\includegraphics[width=\textwidth]{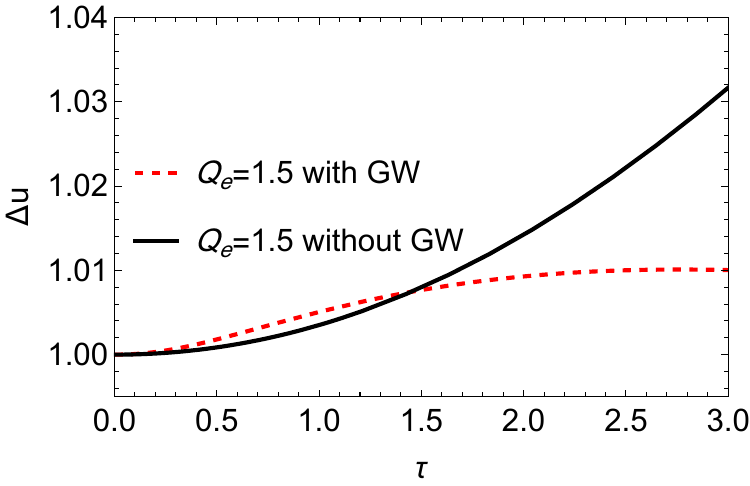}
		\caption{Variation of $\Delta u$ due to GW pulse and without GW pulse with respect to $\tau$ of TWH.( $\mathcal{M}=1$, $A_{0}=1$ and $r_{0}=2$)}
		\label{fig21}
	\end{minipage}
	\hfill
	\begin{minipage}[b]{0.4\textwidth}
		\includegraphics[width=\textwidth]{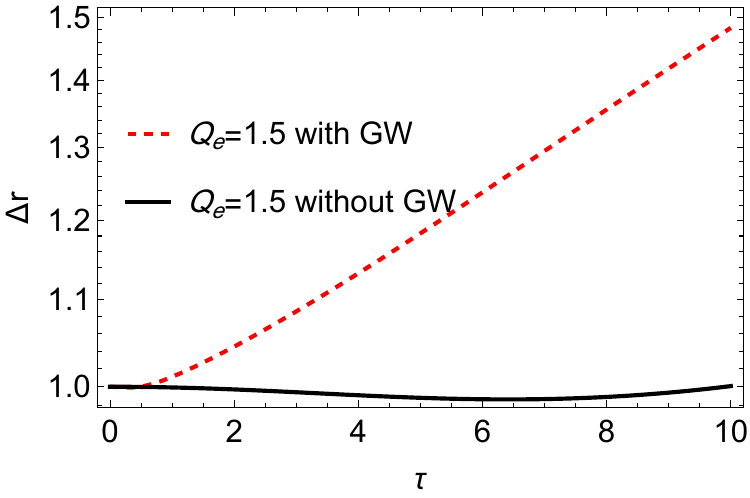}
		\caption{ Variation of $\Delta r$ due to GW pulse and without GW pulse with respect to $\tau$ of TWH.( $\mathcal{M}=1$, $A_{0}=1$ and $r_{0}=2$)}
		\label{fig22}
	\end{minipage}
\end{figure}
\begin{figure}[H]
	\begin{minipage}[b]{0.4\textwidth}
		\includegraphics[width=\textwidth]{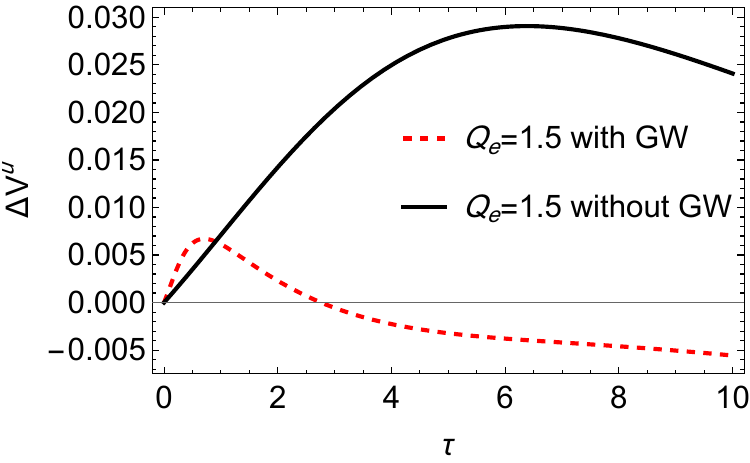}
		\caption{Variation of $\Delta V^{u}$ due to GW pulse and without GW pulse with respect to $\tau$ of TWH.( $\mathcal{M}=1$, $A_{0}=1$ and $r_{0}=2$)}
		\label{fig23}
	\end{minipage}
	\hfill
	\begin{minipage}[b]{0.4\textwidth}
		\includegraphics[width=\textwidth]{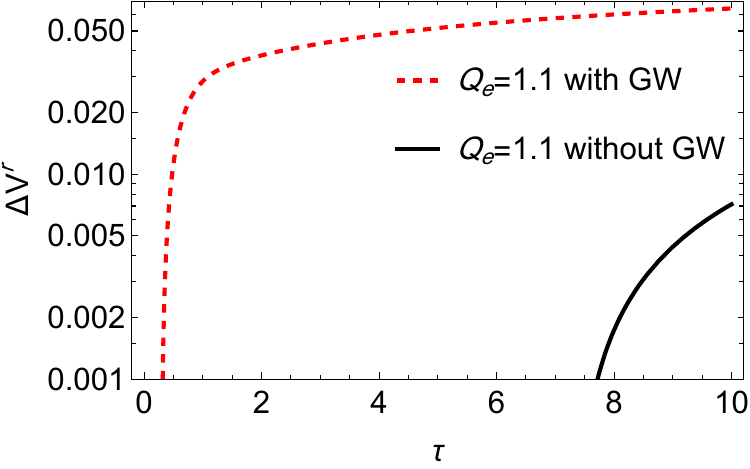}
		\caption{ Variation of $\Delta V^{r}$ due to GW pulse and without GW pulse with respect to $\tau$ of TWH. $\mathcal{M}=1$, $A_{0}=1$ and $r_{0}=2$)}
		\label{fig24}
	\end{minipage}
\end{figure}

Figures \ref{fig21} to \ref{fig24} provide a clear demonstration of the tangible GW pulse within the context of the traversable wormhole (TWH) for certain values of electric charge $Q_{e}$ and parameters $\mathcal{M}=1$, $A_{0}=1$ and $r_{0}=2$. These figures showcase both the displacement and velocity memory effects. Specifically, figures \ref{fig21} and \ref{fig22} depict the changes in displacement memory effects when the GW pulse is either present or absent for an electric charge of $Q_{e}=1.5$. Conversely, figures \ref{fig23} and \ref{fig24} illustrate the variations in $\Delta V^{u}$ and $\Delta V^{r}$ in the presence or absence of the GW pulse, once again for an electric charge of $Q_{e}=1.5$ for $\Delta V^{u}$ and $Q_{e}=1.1$ for $\Delta V^{r}$ . The geodesic deviation of two neighboring points decreases for the $u$ component of displacement and velocity memory effects when the GW pulse is present. However, this physical phenomenon is reversed for the $r$ component.

\section{Black bounce solution and Bondi-Sachs formalism}
Within this segment, we provide a concise overview of the Bondi-Sachs formalism and examine how the presence of a black bounce background and GWs impacts the change in Bondi mass. The Simpson-Visser and Bardeen-Type solutions for black bounce are explored in the subsequent sections while considering the influence of GWs.
\subsection{Simpson-Visser solution and Bondi-Sachs formalism}
The investigation of gravitational memory effects when a GW pulse propagates on the background metric $(\ref{metricRR})$ is conducted using the Bondi-Sachs formalism at null infinity. The gravitational field equations $G_{\mu\nu}=8\pi G \mathcal{T}_{\mu\nu}$ are solved, where the effective energy momentum tensor $\mathcal{T}_{\mu\nu}$ can include scalar field, linear or non-linear electromagnetic fields, or other appropriate fields \cite{ata,silva}. These equations are similar to the generalized GR equations, as shown in \cite{bms}. By employing the Bondi-Sachs method, the impact of the GW pulse on the Bondi mass as it approaches null infinity is analyzed. It is worth noting that the background metric $(\ref{metricRR})$ remains static and does not exhibit any variations in Bondi mass in the absence of the GW pulse.

 The metric $(\ref{metricRR})$ exhibits asymptotic behavior that can be described by the following expression:
 \begin{equation}\label{vissermetric}
     ds^{2}=-(1-\frac{2m}{R})du^{2}+(-2-\frac{a^{2}}{R^{2}})dudR+R^{2}(d\theta^{2}+\sin{\theta}^{2}d\phi^{2}).
 \end{equation}
 
To achieve our objective, we impose the GW components on the null coordinate metric $(\ref{metricRR})$ in the Bondi-Sachs form. This results in the following perturbation metric:
 \begin{eqnarray}\label{perturbationmetric}
 ds^{2}&=&-du^{2}-2dudR+(\frac{2m}{R}+\frac{2M_{B}(u,\theta)}{R}+\frac{E(u,\theta)}{R^{2}})du^{2}-\\\nonumber
 && 2(\frac{a^{2}}{2R^{2}}
 +\frac{b_{1}(u,\theta)}{R}+\frac{b_{2}(u,\theta)}{R^{2}})dudR
 -2R^{2}U(u,R,\theta)dud\theta+R^{2}h_{AB}dx^{A}dx^{B}....,
 \end{eqnarray}
  where $h_{AB}$ is given by 
 \begin{equation}
 	h_{AB}=\gamma_{AB}+C_{AB}R^{-1}+D_{AB}R^{-2}+\mathcal{O}(R^{-3}),
 \end{equation} 
In this expression, the shear tensor $C_{AB}$ is denoted by $C_{AB}=C_{AB}(u,x^{A})$, where $A , B=1,2$. Additionally, the leading order tensor induced by the GW pulse is represented by $D_{AB}=D_{AB}(u,x^{A})$. However, for the purpose of our analysis, this term is disregarded. The GW pulse affects the dynamic Bondi mass $M_{B}(u,\theta)$. Furthermore, the metric $(\ref{perturbationmetric})$ includes other dynamical terms that are perturbations of the background metric caused by the GW pulse.
 
We utilize the Bondi-Sachs metric for the resolution of the gravitational field equation in a perturbative manner. The off-shell solution of the Bondi metric is introduced, characterized by the line element given by 
 \begin{equation}\label{bondi-sachs}
 	ds^{2}=-e^{2\beta}Fdu^{2}-2e^{2\beta}dudR+R^{2}h_{AB}(dx^{A}-\frac{U^{A}}{R^{2}}du)(dx^{B}-\frac{U^{B}}{R^{2}}du).
 \end{equation}
 In order to address the gravitational field equation perturbatively, we substitute the metric $(\ref{bondi-sachs})$ into the Einstein field equation $G_{\mu\nu}=8\pi G \mathcal{T}_{\mu\nu}$ and focus on the large $R$ limit at null infinity. Subsequently, we compare the resolved metric with the metric $(\ref{perturbationmetric})$ term by term to determine the variation in the Bondi mass $M_{B}(u,\theta)$. The forthcoming analysis will present this comparison, preceded by an explanation of certain aspects of the Bondi shear tensor and the News tensor.

The GW pulse in our scenario is characterized by the tensorial News $N_{AB}$, which is expressed as
 \begin{equation}\label{c}
    N_{AB} =\partial_{u}C_{AB}= A_{0}~\text{sech}^{2}(u) Y 
     \begin{pmatrix}
     1&0\\
     0&-\sin^{2}\theta
     \end{pmatrix}.
 \end{equation}
In our scenario, the GW pulse is defined by the News tensor $N_{AB}$, which is determined by the $\text{sech}^{2}(u)$ function. This specific choice is in line with the upcoming section and signifies the impact of the GW pulse on the background geometry. The News tensor $N_{AB}$ represents the gravitational degrees of freedom. To ensure that the GW pulse has a minimal effect on the hairy black hole solution $(\ref{metricRR})$, we adjust the value of $A_{0}$. Subsequently, we utilize equation $(\ref{c})$ to compute the shear tensor $C_{AB}$,
\begin{equation}
     C_{AB}=A_{0}Y
     \begin{pmatrix}
          1&0\\
     0&-\sin^{2}\theta
     \end{pmatrix}\int_{-\infty}^{+\infty}\text{sech}^{2}(u)du=-2Y\begin{pmatrix}
          A_{0}&0\\
     0&-A_{0}\sin^{2}\theta
     \end{pmatrix},
 \end{equation}
 where $Y$ is the spin-weighted harmonic which is defined by $Y=\frac{3}{4}\sqrt{\frac{5}{6\pi}}\sin^{2}\theta$.
 
 By using the metric $(\ref{bondi-sachs})$, one has the ability to utilize it in solving the gravitational field equations in order to obtain the consequent results from the effective Einstein field equations. The hypersurface equation $E_{\mu\nu}=R_{\mu\nu}-\frac{1}{2}g_{\mu\nu}R-8\pi G_{0}\bar{T}_{\mu\nu}=0$ gives us the equation $E^{u}_{R}=0$, leading to
\begin{equation}
\partial_{R}\beta=\frac{R}{16}h^{AC}h^{BD}\partial_{R}h_{AB}\partial_{R}h_{CD}+2\pi \bar{T}_{RR}.
\end{equation}

In the context of the Bondi-Sachs null coordinate system, the source $\bar{T}_{RR}$ is reduced to zero for the particular scenario being examined. As a result, the value of $\beta$ can be determined in the following manner. 
\begin{equation}\label{c1}
\beta= \frac{\beta_{2}}{R^{2}}+\mathcal{O}(R^{-3}),
\end{equation}
Here, $\beta_{2}$ is defined as $-C^{2}/32$ and $C^{2}$ is equal to $C_{AB}C^{AB}$. To determine $U^{A}$ and $F$, we utilize the hypersurfaces $E^{u}_{R}=0$ and $E_{u}^{u}=0$, respectively \cite{bms,772}.

Therefore, considering the factors mentioned above and expanding the function $F$ to the first order, we can derive the following equation:
 \begin{equation}\label{c2}
 	F=1-\frac{2M}{R}-\frac{2F_{2}}{R^{2}}+\mathcal{O}(R^{-3}).
 \end{equation}
The determination of the Bondi mass, represented as $M$, can be achieved through the balance equation, which will be derived in this section. The detailed calculation can be found in reference\cite{772}. By utilizing the condition $E^{u}_{R}=0$, it can be inferred that \cite{772},
 \begin{equation}\label{c3}
 	U^{A}=U^{0A}+\frac{\sigma(u,x^{A})}{R}+\mathcal{O}(R^{-2}),
 \end{equation} 
the term $ \sigma^{A} \equiv\sigma(u,x^{A})$ is obtained by
\begin{equation}
    \sigma^{A}=-\frac{2}{3}L^{A}+\frac{1}{16}D^{A}(C_{BC}C^{BC})+\frac{1}{2}C^{AB}D^{C}C_{BC}.
\end{equation}
The angular momentum aspect $L^{A}(u,\theta)$ is defined as the derivative of it with respect to $u$ given by
\begin{equation}
    \dot{L}_{A}=D_{A}M+\frac{1}{2}D^{B}D_{[A}D^{C}C_{B]C}+\frac{1}{4}D_{B}(N^{BC}C_{AC})+\frac{1}{2}D_{B}N^{BC}C_{AC}.
\end{equation}
Here, $D_{A}$ represents the covariant derivative with respect to $\gamma_{AB}$ (spherical Levi-Civita connection) and $A$, $B$ are raised and lowered by the static metric $\gamma_{AB}$. It is important to note that the densities of energy and angular momentum on celestial spheres are determined by the Bondi mass $M$ and the angular momentum aspect $L_{A}(u,\theta)$.

Ultimately, by using equations $(\ref{c1})$, $(\ref{c2})$ and $(\ref{c3})$,  the on-shell metric is given by
 \begin{eqnarray}\label{16a}
 	ds^{2}&=&-(1-\frac{2M}{R}-\frac{2F_{2}}{R^{2}})du^{2}-2(1-\frac{\beta_{2}}{R^{2}})dudR+	(R^{2}\gamma_{AB}+RC_{AB}+D_{AB})dx^{A}dx^{B}\\\nonumber
 &&	-2(\gamma_{AB}U^{0A}+\frac{\gamma_{AB}\sigma^{A}+C_{AB}U^{0A}}{R}+\frac{C_{AB}\sigma^{A}+D_{AB}U^{0A}}{R^{2}})dudx^{B}\\\nonumber
 	&&+\gamma_{AB}\frac{U^{0A}U^{0B}}{R^{2}}du^{2}+\mathcal{O}(R^{-3}).
 \end{eqnarray}
By comparing $(\ref{16a})$ with the metric $(\ref{perturbationmetric})$ one can obtain
 \begin{align}\label{compare1}
 	&~~~~~~~~~~~~~~~~~~~~~~~~~~~~~~~~~~~~~~~M_{B}(u,\theta)=M-m,&\\
\label{compare2}
 &~~~~~~~~~~~~~~~~~E(u,\theta)=-2(\frac{1}{6}(D_{A}L^{A})+\frac{1}{8}(D_{A}C^{AB})(D_{D}C_{B}^{D}))-\frac{1}{16}C^{2},&\\
\label{compare3}
 &~~~~~~~~~~~~~~~~~~~~~~b_{2}(u,\theta)=\frac{1}{32}C^{2}-\frac{a^{2}}{2} ~~~~~~~,~~~~~~C^{2}=C_{AB}C^{AB},&\\
 \label{compare4}
 &~~~~~~~~~~~~~~~~~~~~~~~~~~~~~~~~~~~~~~~~~~~~~~b_{1}(u,\theta)=0,&\\
 \label{compare5}
 &U(u,R,\theta)=\frac{\gamma_{AB}U^{0A}}{R^{2}}+\frac{\gamma_{AB}(-\frac{2}{3}L^{A}+\frac{1}{16}D^{A}C^{2}+\frac{1}{2}C^{AB}D^{C}C_{BC})+C_{AB}U^{0A}}{R^{3}}&\\\nonumber
 &~~~~~~~~~~~~~~~~~+\frac{C_{AB}(-\frac{2}{3}L^{A}+\frac{1}{16}D^{A}C^{2}+\frac{1}{2}C^{AB}D^{C}C_{BC})+D_{AB}U^{0A}}{R^{4}}.&
 \end{align}
In equation $(\ref{compare1})$, $M$ represents the Bondi mass and is determined by the balance equation \cite{772} given as
 \begin{equation}\label{m}
     \dot{M}=\frac{1}{4}D_{A}D_{B}N^{AB}-\frac{1}{8}N_{AB}N^{AB}-4\pi R^{2}\bar{T}_{uu},
 \end{equation}
where the dot denotes differentiation with respect to $u$. For the Kiselev solution and its asymptotic behavior, $\bar{T}_{uu}=0$. By substituting the News tensor from equation $(\ref{c})$ into $(\ref{m})$, one can derive
%\begin{figure}
\begin{figure}[H]
    \centering
    \includegraphics[width=0.5\linewidth]{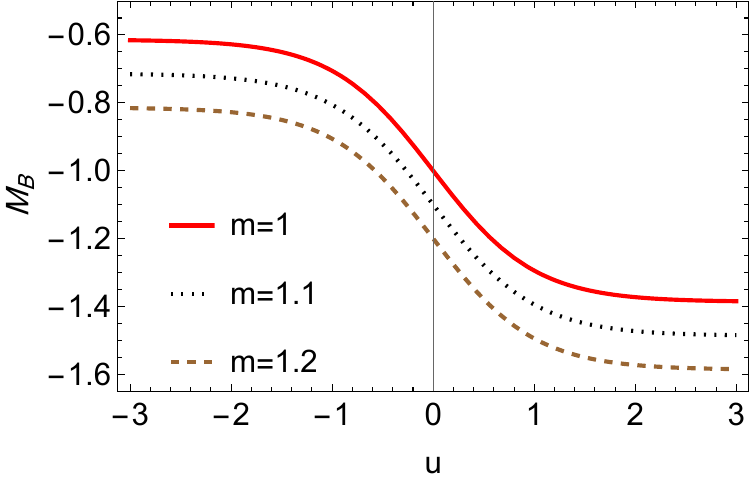}
    \caption{Variation in the change of the Bondi mass for Simpson-Visser solution for $A_{0}=2$ and for different values of mass $m$.}
    \label{fig:Bondiblack}
\end{figure}

\begin{equation}\label{mm}
   \dot{M}=\partial_{u}M= \frac{1}{4}A_{0}~\text{sech}^{2}u(Y_{\theta \theta}+Y\cos{2\theta}-Y_{\theta}\cot{\theta}+2Y\cot^{2}\theta) -\frac{1}{4}A_{0}^{2}Y^{2}\text{sech}^{4}u,
\end{equation}
$Y_{\theta}$ and $Y_{\theta \theta}$ represent the first and second derivatives of $Y$ with respect to $\theta$, respectively. By referring to equation $(\ref{mm})$, we can obtain $M$
\begin{equation}\label{balance}
    M= \frac{1}{4}A_{0}(Y_{\theta \theta}+Y\cos{2\theta}-Y_{\theta}\cot{\theta}+2Y\cot^{2}\theta)\int  \text{sech}^{2}udu-\frac{1}{4}A_{0}^{2}Y^{2}\int  \text{sech}^{4}udu.
\end{equation}
By manipulating the equation $(\ref{balance})$ we have
\begin{equation}\label{M1}
    M= \frac{1}{4}A_{0}\tanh{u}(Y_{\theta \theta}+Y\cos{2\theta}-Y_{\theta}\cot{\theta}+2Y\cot^{2}\theta)-\frac{1}{4}A_{0}^{2}Y^{2}\tanh{u}(\frac{2}{3}+\frac{1}{3}\text{sech}^{2}u).
\end{equation}

Therefore, by applying equations $(\ref{compare1})$ and $(\ref{M1})$, one can obtain the change in Bondi mass as follows
 \begin{equation}\label{Mbondi}
    M_{B}(u,\theta)=-m + \frac{1}{4}A_{0}\tanh{u}(Y_{\theta \theta}+Y\cos{2\theta}-Y_{\theta}\cot{\theta}+2Y\cot^{2}\theta)-\frac{1}{4}A_{0}^{2}Y^{2}\tanh{u}(\frac{2}{3}+\frac{1}{3}\text{sech}^{2}u).
 \end{equation}
The Bondi mass transformation with fluctuations in the black bounce mass $m$ is demonstrated in Figure \ref{fig:Bondiblack}. It is specifically shown for $M_{B}$ versus $u$ on the $\theta=\pi/2$ plane, leading to the equation for $M_{B}$. Therefore, the Bondi mass is given by
 \begin{equation}\label{bondialtered}
     M_{B}(u,\theta=\frac{\pi}{2})=-m-\frac{3}{8}\sqrt{\frac{5}{6 \pi}} A_{0}\tanh(u).
 \end{equation}
 The diagram presented in Figure \ref{fig:Bondiblack} showcases the variation in the Bondi mass at null infinity for large values of $R$ caused by GWs in the Simpson-Visser background. The value of $M_{B}$ is calculated for $A_{0}=2$ and various values of $m$, as depicted in Figure \ref{fig:Bondiblack}. With an increase in $m$, the magnitude of the alteration in Bondi mass also increases. Consequently, the disturbance of the gravitational central object can be observed at future null infinity by monitoring the changes in Bondi mass.

\subsection{Bardeen-Type black bounce and Bondi-Sachs formalism}
In this section, we will examine the impact of GWs on the background of a Bardeen-Type black bounce on the change of Bondi mass at null infinity. Consequently, we will reiterate subsection (A) to analyze the Bardeen-Type solution of the black bounce. To achieve this, we will rewrite the metric $(\ref{metricRR})$ by employing the relevant equations $(\ref{F})$ and $(\ref{G})$  as follows 
\begin{equation}\label{RR}
     ds^{2}=-(1-\frac{2m}{R}+\frac{2ma^{2}}{R^{3}})du^{2}-2(1-\frac{a^{2}}{R^{2}})^{-1/2}dudR+R^{2}(d\theta^{2}+\sin{\theta}^{2}d\phi^{2}).
 \end{equation}
  The asymptotic behaviour of metric  $(\ref{RR})$ up to order $R^{-2}$ can be given by
 \begin{equation}\label{bardeenmetric}
     ds^{2}=-(1-\frac{2m}{R})du^{2}+(-2-\frac{a^{2}}{R^{2}})dudR+R^{2}(d\theta^{2}+\sin{\theta}^{2}d\phi^{2}),
 \end{equation}
 Which is equivalent to metric $(\ref{vissermetric})$. Upon comparing $(\ref{16a})$ with the metric $(\ref{bardeenmetric})$, we derive equations $(\ref{compare1})$ to $(\ref{compare5})$ for the Bardeen-Type solution, similar to what we have for the Simpson-Visser black bounce. Consequently, by examining the change in Bondi mass at future null infinity for large $R$, distinguishing between the Bardeen-Type solution and the Simpson-Visser black bounce becomes challenging.
 \section{TRAVERSABLE WORMHOLE SOLUTION and Bondi-Sachs formalism}
 In this part, we will replicate the identical approach utilized in section V for a traversable wormhole solution to analyze the change in Bondi mass with respect to the parameter of this particular wormhole solution.

To achieve this, it is essential to establish the metric $(\ref{tortoiseWH})$ at the primary level to comprehend its characteristics over significant distances. Hence, we obtain.
\begin{eqnarray}\label{mTWH}
    ds^{2}=-du^{2} +(\frac{2\mathcal{M}}{r}-\frac{\mathcal{M}^{2}}{r^{2}})du^{2}
    -2dudr-2(\frac{\frac{Q_{e}^{2}}{2r_{0}}+\frac{r_{0}}{2}-\mathcal{M}}{r}+\frac{-\frac{Q_{e}^{2}}{4}+\frac{3Q_{e}^{4}}{8r_{0}^{2}}+\frac{3r_{0}^{2}}{8}-\frac{\mathcal{M}Q_{e}^{2}}{2r_{0}}-\frac{\mathcal{M}r_{0}}{2}}{r^{2}})dudr+r^{2}d\Omega^{2}.
\end{eqnarray}
Now, we disturb the background metric $(\ref{mTWH})$ by introducing a GW pulse. The perturbed metric is given by:
\begin{eqnarray}\label{perturbWH}
  ds^{2}&=&-du^{2} +(\frac{2\mathcal{M}}{r}+\frac{2M_{B}(u,\theta)}{r}-\frac{\mathcal{M}^{2}}{r^{2}}+\frac{E(u,\theta)}{r^{2}})du^{2}
    -2dudr\\\nonumber
    &&-2(\frac{\frac{Q_{e}^{2}}{2r_{0}}+\frac{r_{0}}{2}-\mathcal{M}}{r}+\frac{b_{1}(u,\theta)}{r}+\frac{-\frac{Q_{e}^{2}}{4}+\frac{3Q_{e}^{4}}{8r_{0}^{2}}+\frac{3r_{0}^{2}}{8}-\frac{\mathcal{M}Q_{e}^{2}}{2r_{0}}-\frac{\mathcal{M}r_{0}}{2}}{r^{2}}+\frac{b_{2}(u,\theta)}{r^{2}})dudr\\\nonumber
    &&-2r^{2}U(u,r,\theta)dud\theta+r^{2}h_{AB}dx^{A}dx^{B}....
\end{eqnarray}
By comparing equation $(\ref{16a})$ with the metric $(\ref{perturbWH})$ and substituting $R$ with $r$ in equation $(\ref{16a})$, we can derive the following relationships,
%\begin{figure}
\begin{figure}[H]
    \centering
    \includegraphics[width=0.5\linewidth]{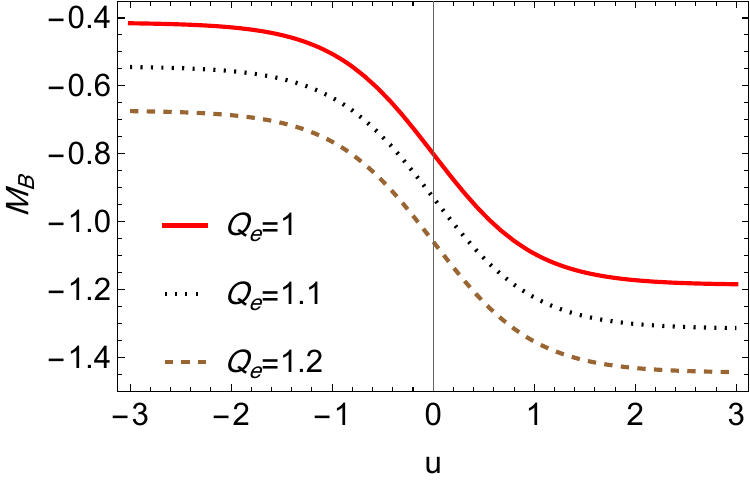}
    \caption{Variation the change of Bondi mass for a traversable wormhole for $A_{0}=2$, $r_{0}=2$ and different values of electric charge $Q_{e}$}
    \label{fig:BondiWHmass}
\end{figure}

\begin{align}\label{comparee1}
 	&~~~~~~~~~~~~~~~~~~~~~~~~~~~~~~~~~~~~~~~~~~M_{B}(u,\theta)=M-\mathcal{M},&\\
\label{comparee2}
 &~~~~~~~~~~~~~~~~~~~~~E(u,\theta)=-2(\frac{1}{6}(D_{A}L^{A})+\frac{1}{8}(D_{A}C^{AB})(D_{D}C_{B}^{D}))-\frac{1}{16}C^{2},&\\
 \label{comparee3}
& ~~~~~~~~~~~~b_{2}(u,\theta)=\frac{1}{32}C^{2}+\frac{Q_{e}^{2}}{4}-\frac{3Q_{e}^{4}}{8r_{0}^{2}}-\frac{3r_{0}^{2}}{8}+\frac{\mathcal{M}Q_{e}^{2}}{2r_{0}}+\frac{\mathcal{M}r_{0}}{2} ,~~~~~~C^{2}=C_{AB}C^{AB},&\\
 \label{comparee4}
 &~~~~~~~~~~~~~~~~~~~~~~~~~~~~~~~~~~~~~~~b_{1}(u,\theta)=-\frac{Q_{e}^{2}}{2r_{0}}-\frac{r_{0}}{2}+\mathcal{M},&\\
 &\label{comparee5}
 ~~~~~~~~~U(u,R,\theta)=\frac{\gamma_{AB}U^{0A}}{r^{2}}+\frac{\gamma_{AB}(-\frac{2}{3}L^{A}+\frac{1}{16}D^{A}C^{2}+\frac{1}{2}C^{AB}D^{C}C_{BC})+C_{AB}U^{0A}}{r^{3}}&\\\nonumber
 &~~~~~~~~~~~~~~~~~~~~~~~~~+\frac{C_{AB}(-\frac{2}{3}L^{A}+\frac{1}{16}D^{A}C^{2}+\frac{1}{2}C^{AB}D^{C}C_{BC})+D_{AB}U^{0A}}{r^{4}}.&
 \end{align}
 Now we use $ \mathcal{M}=\frac{2Q_{e}^{2}r_{0}}{Q_{e}^{2}+r_{0}^{2}}$ in equation $(\ref{comparee1})$ to obtain
 \begin{equation}\label{compareee}
 	M_{B}(u,\theta)=M-\frac{2Q_{e}^{2}r_{0}}{Q_{e}^{2}+r_{0}^{2}}.
 \end{equation}
Figure \ref{fig:BondiWHmass} illustrates the change in the Bondi mass as the traversable wormhole with charge $Q_{e}$ varies. The graph shows the relationship between $M_{B}$ and $u$ for the $\theta=\pi/2$ plane. The equation for $M_{B}$ in this case is given by
 \begin{equation}\label{bondialteredWH}
     M_{B}(u,\theta=\frac{\pi}{2})=-\frac{2Q_{e}^{2}r_{0}}{Q_{e}^{2}+r_{0}^{2}}-\frac{3}{8}\sqrt{\frac{5}{6 \pi}} A_{0}\tanh(u),
 \end{equation}
which is derived using equation $(\ref{compareee})$ and the change in Bondi mass $(\ref{Mbondi})$. It is important to note that $M_{B}$ is calculated based on the values of $Q_{e}$ and $r_{0}$, and the term $A_{0}$ is also included in the equation.

The graph shown in Figure \ref{fig:BondiWHmass} illustrates the variation in Bondi mass for different values of electric charge $Q_{e}$, with $A_{0}=2$ and $r_{0}=2$. Its purpose is to observe the impact of the gravitational Pulse on the geometric background of the traversable wormhole. It is evident that the magnitude of the change in Bondi mass becomes greater as the electric charge $Q_{e}$ increases. Consequently, the alteration in the background metric caused by the electric charge can be detected through the change in Bondi mass at null infinity.

\section{Conclusion}
The present study examines the gravitational memory effects, specifically the displacement and velocity memories, in the context of the black bounce and EDM traversable wormhole solutions. The investigation considers both in the presence and absence of a GW pulse. The findings reveal that in the case of the Simpson-Visser and Bardeen-Type black bounce solutions, the presence of a GW pulse leads to noticeable changes in the two neighboring geodesics. This phenomenon is also observed in the TWH solution, which incorporates fermion and Maxwell fields to enable traversability. In summary, the results demonstrate the impact of GW pulses on the behavior of geodesics in these solutions. We summarize the results as follows:

1- Simpson-Visser Solution: As the magnetic charge $a$ increases, the difference between two adjacent geodesics for the component $u$ $(\Delta u)$ decreases, while for the component $R$, the difference increases. These disparities can be interpreted as a displacement memory effect. The velocity memory effect refers to the changes in velocities of $u$ and $R$, determined by the variations of $\Delta V^{u}$ and $\Delta V^{R}$ with respect to $\tau$.

2-Bardeen-Type Black Bounce: Similar to the Simpson-Visser solution, the displacement and velocity memory effects are also applicable. As the magnetic charge a increases, the difference between two adjacent geodesics for the component $u$ $(\Delta u)$ increases, while for the component R, the difference decreases. The changes in $u$ and $R$ can be interpreted as a memory effect related to displacement. The velocity memory effect refers to the variation in the velocities of u and R, determined by the variations of  $\Delta V^{u}$ and $\Delta V^{R}$ with respect to $\tau$.

3-Traversable Wormhole Solution: Similar to the previous solutions for black bounces, the displacement memory effect denoted as $\Delta r$ and $\Delta u$, as well as the velocity memory effects represented by  $\Delta V^{u}$ and $\Delta V^{r}$ are observed. These effects are obtained through numerical solutions of certain equations.

In all cases, the presence or absence of a GW pulse also affects these memory effects. The figures mentioned in the text provide visual demonstrations of these phenomena.

Moreover, this paper investigates the memory effects of GWs. We analyze the black bounce solutions, such as Simpson-Visser and Bardeen-Type, as well as a traversable wormhole solution. These solutions serve as a background that is perturbed by a GW pulse at null infinity and for large values of $r$. To accomplish this, we utilize the Bondi-Sachs formalism to study the memory effect of the black bounces and traversable wormhole at null infinity. We introduce the GW pulse into the background of these solutions to observe the change in Bondi mass. The Bondi mass is a measure of mass on an asymptotically null slice or the densities of energy on celestial spheres. We focus on specific solution parameters, such as the mass $m$ for black bounces and the electric charge $Q_{e}$ for the wormhole, and examine their effects on the change of Bondi mass. The variations in Bondi mass at null infinity are presented in figures  \ref{fig:Bondiblack} and \ref{fig:BondiWHmass}. As the parameters $m$ and $Q_{e}$ increase, the magnitude of the variation in the change of Bondi mass becomes significant, as depicted in figures  \ref{fig:Bondiblack} and \ref{fig:BondiWHmass}. This variation is observed at future null infinity and for substantial $r$, indicating the impact of these parameters and the perturbation effects of the background geometry at a considerable distance from the central objects.

 %%%%%%%%%%%%%%%%%%%%%%%%%%%%%%%%%%%%%%%%%%%%%%%%%%%%%%%
%%%%%%%%%%%%%%%%%%%%%%%%%%%%%%%%%%%%%%%%%%%%%%%%%

\end{document}